\documentclass[twocolumn,preprintnumbers,amsmath,amssymb,prl ]{revtex4-1}

\usepackage{times}

\usepackage{amsmath}
\usepackage{amsfonts}
\usepackage{amssymb}
\usepackage{graphicx}
\usepackage{color}

\newcommand*{\citen}{}
\DeclareRobustCommand*{\citen}[1]{%
  \begingroup
    \romannumeral-`\x 
    \setcitestyle{numbers}%
    \cite{#1}%
 \endgroup
}

\definecolor{mypink}{RGB}{219, 48, 122}
\definecolor{mygreen}{RGB}{51, 153, 102}
\definecolor{brown}{RGB}{165, 42, 42}

\newcommand{\EF}{$\mathrm{E}_\mathrm{F}$}

 \newcommand{\Tc}{$T_\mathrm{C}$ }

\newcommand{\kz}{$k_\mathrm{z}$}
\newcommand{\hn}{$\mathrm{h\nu}$}

 \newcommand{\GM}{$\overline{\Gamma\mathrm{M}}$}

 \newcommand{\GK}{$\overline{\Gamma\mathrm{K}}$}

 \newcommand{\KGK}{$\overline{\mathrm{K}\Gamma\mathrm{K}}$}

\newcommand{\ZU}{$\mathrm{Z}$-$\mathrm{U}$}
\newcommand{\ZA}{$\mathrm{Z}$-$\mathrm{A}$}

\newcommand{\aGeTe}{$\alpha$\nh GeTe}

\newcommand{\GMT}{Ge$_{1-x}$Mn$_x$Te}
\newcommand{\GMTD}{Ge$_{0.87}$Mn$_{0.13}$Te}
\newcommand{\GMTDD}{Ge$_{0.94}$Mn$_{0.06}$Te}

\newcommand{\n}[1]{$n$\nobreakdash-\hspace{0pt}}
\newcommand\nh{\mbox{-}}

\newcommand\dg{$^{\circ}$}

\newcommand{\Pz}{$P_{z}$}
\newcommand{\Pxy}{$P_{x,y}$}
\newcommand{\Px}{$P_{x}$}
\newcommand{\Py}{$P_{y}$}
\newcommand{\ooo}{$\langle 111 \rangle$}
\newcommand{\kA}{$\mbox{\AA}^{-1}$}

\begin{document} 

\title {Spin-resolved electronic structure of ferroelectric \aGeTe\ and multiferroic \GMT}

\author{J.~Krempask{\'y}${}^{1}$, M.~Fanciulli${}^{1,2}$, N.~Pilet${}^{1}, $J.~Min{\'a}r${}^{3}$ , W.~Khan${}^{3}$, M.~Muntwiler${}^{1}$, F.~Bertran${}^{4}$, S.~Muff${}^{1,2}$, A.P.~Weber${}^{1,2}$, V.N.~Strocov${}^{1}$,V.V.~Volobuiev${}^{5,6}$,  G.~Springholz${}^{6}$, J.~H.~Dil${}^{1,2}$
}

\affiliation{
$^1$Swiss Light Source, Paul Scherrer Institut, CH-5232 Villigen PSI, Switzerland\\
$^2$Institute of Physics, \'Ecole Polytechnique F\'ed\'erale de Lausanne, CH-$1015$ Lausanne, Switzerland\\
$^3$New Technologies-Research Center University of West Bohemia, Plze{\v n}, Czech Republic\\
$^4$SOLEIL Synchrotron, L'Orme des Merisiers, F-91192 Gif-sur-Yvette, France\\
$^5$National Technical University, Kharkiv Polytechnic Institute, Frunze Str. 21, 61002 Kharkiv, Ukraine\\
$^6$Institut f{\"u}r Halbleiter-und Festk{\"o}rperphysik, Johannes Kepler Universit{\"a}t, A-4040 Linz, Austria\\
}

\begin{abstract}

Germanium telluride features special spin-electric effects originating from spin-orbit coupling and symmetry breaking by the ferroelectric lattice polarization, which opens up many prospectives for electrically tunable and switchable spin electronic devices. By  Mn doping of the  \aGeTe\ host lattice, the system becomes a multiferroic semiconductor possessing magnetoelectric properties in which the electric polarization, magnetization and spin texture are coupled to each other.  Employing spin- and angle-resolved photoemission spectroscopy in bulk- and surface-sensitive energy ranges and by varying dipole transition matrix elements, we disentangle the bulk, surface and surface-resonance states of the electronic structure and determine the spin textures for selected parameters. From our results we derive a comprehensive model of the \aGeTe\ surface electronic structure which fits to experimental data and first principle theoretical predictions and we discuss the unconventional evolution of the Rashba-type spin splitting upon manipulation by external B- and E-fields.

\end{abstract}

\maketitle

\section{INTRODUCTION} 

Rashba-type effects have been first observed in quantum confined two-dimensional electronic states of semiconductor heterostructures due to the artificial structural asymmetry created at the heterointerfaces\cite{Nitta_PRL, Maekawa_book}. The Rashba splitting of these electronic states can be tuned electrically but the splitting is rather small, limiting practical device applications. In ferroelectrics the large natural structral asymmetry due to the ferroelectric (FE)  lattice displacements leads to a large Rashba splitting even of the bulk bands for which reason such materials have been named ferroelectric Rashba semiconductor (FERS)\cite{Picozzi_AdvM}. The most prominent example is \aGeTe\ featuring a record spin splitting and Rashba parameters \cite{JK_PRB}.   
From the technological point of view GeTe also belongs to a class of chalcogenide phase-change materials \cite{Kolobov_PRB,Dronskowski} and it is the ferroelectric semiconductor with the simplest conceivable binary structure {\cite{Esaki66,Pawley_1966} with strongly asymmetric arrangement of the Ge and Te atoms along the \ooo\ direction \cite{JK_PRB}.

\begin{figure}[t!]
\centering
\includegraphics[width=7cm]{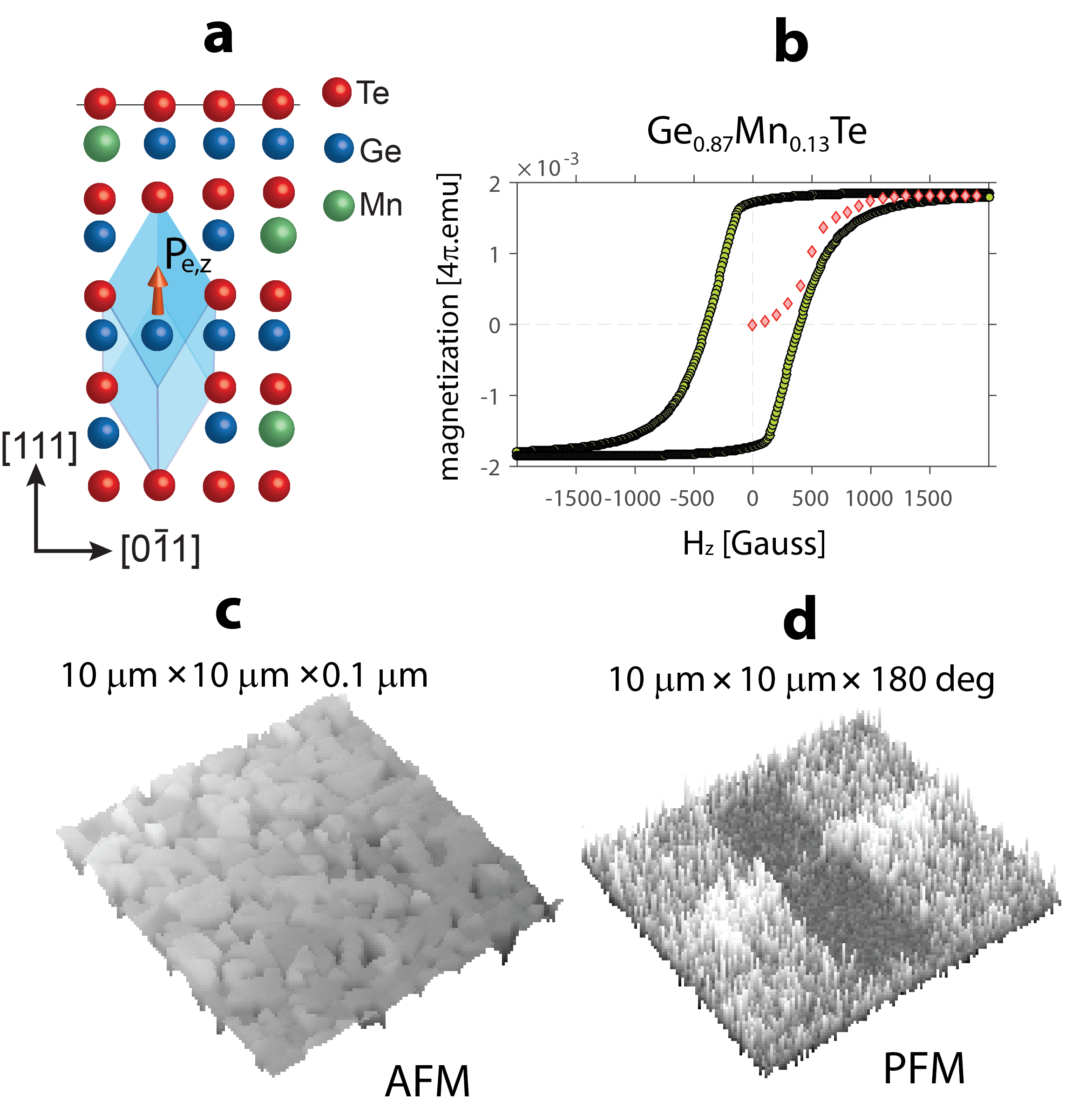}
\caption{(a) Sketch of  multiferroic \GMT\ with ferroelectric displacement of Ge(Mn)-atoms inside the rhombohedrally distorted unit cell along [111] as indicated by the orange arrow. (b) Out-of-plane ferromagnetic hysteresis curve of multiferroic \GMTD\ measured by SQUID. (c) Surface topography measured in atomic force microscopy (AFM). (d) Piezo-force microscopy (PFM) showing 180\dg\ phase change in writing domains forming a cross. The experimental setup is measuring AFM and PFM data simultaneously.} 
\label{F0}
\end{figure}

\begin{figure*}[t]
\centering
\includegraphics[width=0.9\textwidth]{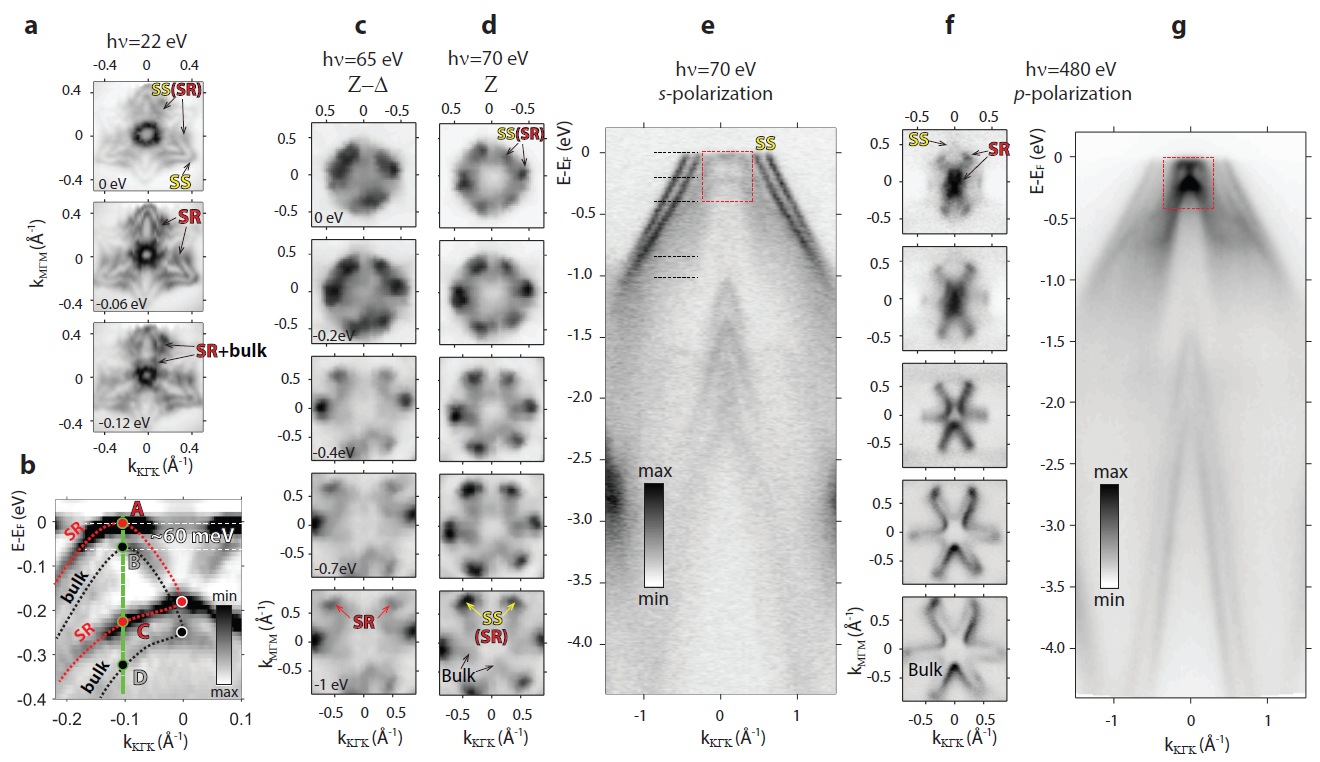}
\caption{ARPES band maps and isosurfaces near the Z\nh point of \aGeTe\} measured at different photon energies: (a,b) \hn=22 eV; (c-e) 65 and 70 eV; (f,g) 480 eV. ARPES data in (c-e) were measured with $s$-type light polarization, remaining data with $p$-type polarization. The isoenergy surfaces (isosurfaces) in panels (c,d,f) were measured at binding energies indicated in panel (e) by horizontal dashed lines.
The arrows indicate momenta for pure surface states (SS) and surface resonances (SR) with respect to bulk bands, also indicated in red dashed rectangles in (e) and (g). Panel (b) is a second derivative band map with bulk bands screened by their resonance replica, vertical green line indicate an EDC-cut intersecting the bands in points A-B-C-D.}
\label{F1}
\end{figure*}

Recently, \aGeTe\ has attracted a flurry of experimental activity \cite{Liebmann_GeTe, JK_PRB, JK_GMT,Schoenhense_GeTe, JK_opGeTe, Bertacco_GeTe} because of its giant Rashba effect, theoretically predicted by S. Picozzi {\cite{Picozzi_AdvM, Picozzi_Front}.
The highly non-centrosymmetric arrangement of the Ge and Te atoms along the \ooo\ direction combined with the large spin-orbit coupling is at the heart of this effect, resulting in the highest reported bulk Rashba coupling parameter $\alpha_R$ of 4.25 eV\AA~\cite{JK_PRB}. Doping of GeTe with Mn leads to additional ferromagnetic (FM) coupling leading to multiferroicity in \GMT\ already for for moderate Mn doping \cite{JK_GMT}. Ferroelectricity is induced by the lattice distortion of GeTe and ferromagnetism by the coupling of the local spins of the Mn ions via the free carriers in the system\cite{Springholz_PRL}. Figure \ref{F0} summarize the \GMT\ thin film basic properties in terms of atomic arrangement (panel a), ferromagnetic hysteresis (panel b), surface topography (panel c) and ferroelectric response measured in piezo-force microscopy (panel d). Due to high Mn solubility and high hole concentration, the FM Curie temperatures of \hbox{\Tc=190~K} is amongst the highest of all FM semiconductors. This new class of materials, termed multiferroic Rashba semiconductors (MUFERS), also display a new type of magnetoelectric switching due to entangled Rashba-Zeeman splitting \cite{JK_GMT}.

Nonetheless, this overwhelming panel of physical properties might also hide unconventional pairings because the system naturally possesses bulk type-II superconductivity in a non-centrosymmetric lattice arrangement \cite{Stiles_GeTe_SC66, GeTe_PSSR}. For this reason further experimental effort is made to engineer topologically non-trivial systems based on \GMT\ by adequate doping in order to optimize material conditions for hosting 'Majorana'-like quasiparticles \cite{Beenakker:2013}.

In this paper, we present a  review of the \aGeTe\ and \GMT\ surface electronic structure studied by (spin- and) angle-resolved photoemission ((S)ARPES). The first issue we address is to show that \aGeTe\ is a narrow gap semiconductor in which the bulk bands, buried inside the \aGeTe\ surface electronic structure probed by ARPES, do not reach the Fermi level in contrast to what was recently claimed \cite{Liebmann_GeTe}. Because there is a conspicuous difference in the ARPES interpretation in this respect, we here demonstrate that pure surface  and bulk states can be clearly distinguished in ARPES and that the surface and bulk Dirac points are well separated in energy. Since surface effects are quenched on capped \aGeTe\  surfaces \cite{JK_PRB}, a direct inspection of bulk states is possible, proving that \aGeTe\ is a ferroelectric Rashba semiconductor with a band gap of about 60 meV. On the other hand, for uncapped surfaces the bulk band edges are difficult to observe due to the presence of strong surface resonance states. This is especially the case near the Z-point where the band gap is smallest and the Rashba splitting is most pronounced \cite{Picozzi_AdvM,JK_PRB}. 
Adding Mn induces ferromagnetism in \aGeTe\, rendering \GMT\ multiferroic at sufficiently low temperature with the magnetization perpendicular to the surface. This ferromagnetism opens a Zeeman gap in the Rashba bands. As shown by SARPES, this moreover leads to a vertical spin polarization at the  Z\nh point of the Brillouin zonen that can be switched by reversal of the magnetization. Magneto-electric coupling further enhances the functionality of which the prospects are discussed.  }

\begin{figure*}[t!]
\centering
\includegraphics[width=.9\textwidth]{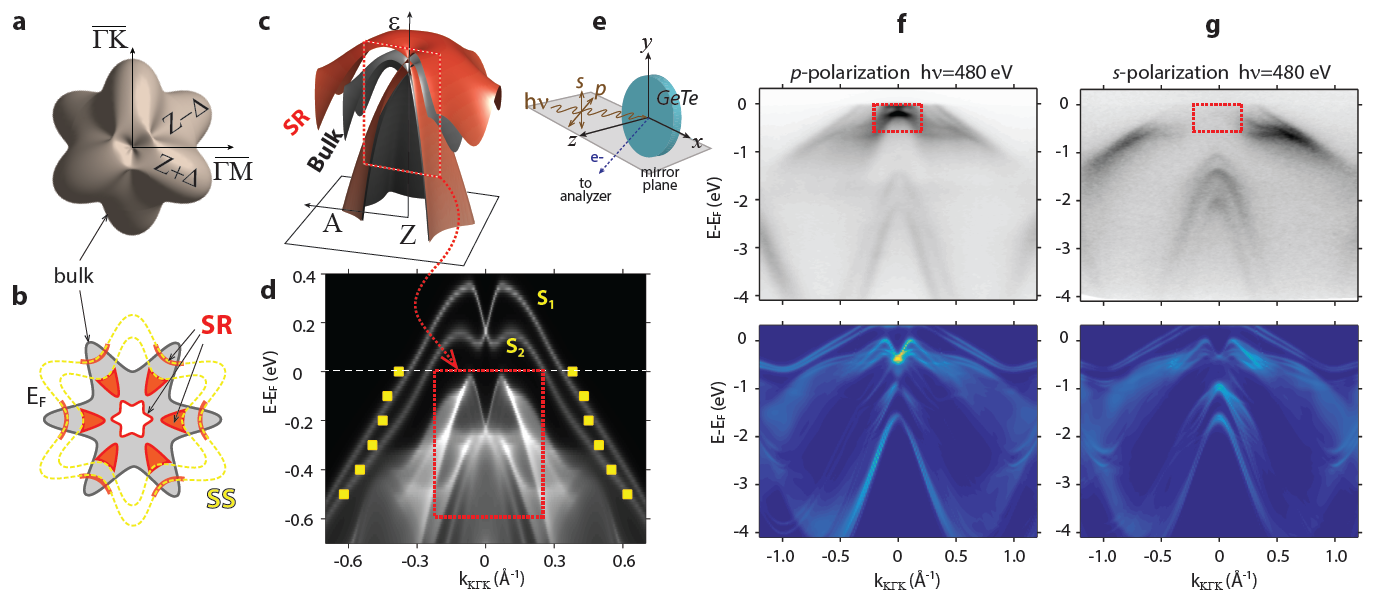}
\caption{(a) 3D schematic representation of the \aGeTe\ bulk isoenergy surface at the Z\nh point and its vicinity (Z$\pm\Delta$). (b) Schematic projection of the \aGeTe\ electronic structure onto the surface Brillouin zone; (c) corresponding  model of the bulk (black) and bulk-derived SR band (red). (d) \aGeTe\ semi-infinite crystal calculations, yellow markers indicate the experimentally retrieved dispersion of the surface states $S_{1,2}$ from Fig.~\ref{F1}d. (e) Experimental geometry with $p$-polarized light. (f,g) ARPES band maps along \KGK\ measured with $p$ and $s$-polarized light, respectively. Red frame indicate the energy and momenta with bulk properties, bottom panels are calculations.} 
\label{F2}
\end{figure*}

\section{$\alpha$-G\MakeLowercase{e}T\MakeLowercase{e} surface electronic structure}

To distinguish the surface electronic structure of \aGeTe\ we compare in Figure~\ref{F1}
ARPES data measured near Z-points  with different photon energies of \hn\ of 22, 70 and 480 eV  at the COPHEE, Pearl and ADRESS photoemission experimental stations at the Swiss Light Source, respectively.
All data were measured at or below 35 K. For each photon energy, constant energy cuts at a given binding energy (isosurfaces) are compared.  The isosurfaces at the Z-point in panels a,d and e of Figure~\ref{F1} have six-fold symmetry, whereas away from the Z\nh points the isosurfaces assume a three-fold symmetry as seen in panel (c). The schematic picture in Fig.~\ref{F2}(a) illustrates how the 6-fold symmetry at the Z-point changes to three-fold above ($Z+\Delta$) and below ($Z-\Delta$) the Z-point, by showing the top-view of the 3D spindle-torus constant energy surface of \aGeTe \cite{JK_PRB}.

The ARPES data in Fig.~\ref{F1} shows the influence of the photoelectron escape depth when probing the same electronic structure in surface sensitive vacuum ultraviolet (\hn=22 eV), bulk sensitive soft-X ray (\hn=480 eV) and in-between (\hn=70 eV) \cite{Fadley:2012}. This comparison allows us to identify the surface states (SS), bulk states and the elusive surface resonances (SR). As extensively discussed in Ref.\citen{JK_PRB}, disentangling the SR and bulk bands for \aGeTe\ near the Z-point is challenging because in the vicinity of the Z-point the SR bands display much higher spectral weight compared to bulk states. Moreover, they disperse with photon energy and are thus easily confused with bulk states \cite{Liebmann_GeTe}. Therefore in ARPES one observes metallic states at \EF, in general agreement with the intrinsic $p$-type doping from Ge vacancies responsible for the metallic character of the nominally semiconducting GeTe \cite{Edwards_PRB}. However, tunnelling experiments provide firm experimental evidence that \aGeTe\ is a narrow-gap semiconductor \cite{Esaki66}. This gap of around 60 meV can also be seen in Fig.~{F1}(b) buried in the surface electronic structure.

Generally speaking, in photoemission experiments the observation of SR bands is expected to occur around the edge of the projected bulk band structure of semiconductors \cite{Spanjaard,Cardona_book,Cardona_PE_book}. In this sense, \aGeTe\ is a textbook example and ignoring the relevance of the SR bands can lead to an erroneous interpretation of the surface electronic structure. This  underlines again the importance to combine bulk and surface sensitive photoemission. The data in Fig.~\ref{F1} reveals the SR-bands detaching from pure surface states in panel (a), progressively enhancing their spectral weight for lower binding energies by forming a 30\dg\ rotated isosurface compared to pure surface states. Their density of states near the next Z-point at \hn=70 eV pile-up at the extremities of the hexagonally-warped bulk states (panels c-d), and for the Z-point probed with \hn=470 eV in panel (f) their spectral weight vanishes because of the increased bulk sensitivity.

Projecting all the isosurfaces on the surface plane from surface- and bulk-sensitive ARPES we see a direct deployment of SR bands detaching from the pure surface states and hybridizing with the bulk continuum, as schematically depicted in Fig.~\ref{F2}b in red. Their isosurface projections at selected binding energies shown in Fig.~\ref{F1}d are overlaid with first-principles calculations (yellow markers in  Fig.~2d) to show that along the mirror planes (in this case along \KGK), the surface resonances follow the dispersion of the two major surface states denoted $S_1$ and $S_2$. We readily see that these surface states have their Dirac point in the unoccupied states because they do not fold back below \EF, and are well separated from the bulk states. In Fig.~\ref{F1}c we observe that SR bands outside the Z-point disperse along the bulk bands by changing the isosurfaces from six to three-fold symmetry, which illustrates how the SR bands mimic the bulk bands, and at the same time, in mirror planes they mimic the surface states. Such observation is typical to surface resonances which materialize in  the sample sub-surface region comparable with photoelectron escape depth (5-10\AA).

\begin{figure}[ht!]
\centering
\includegraphics[width=9cm]{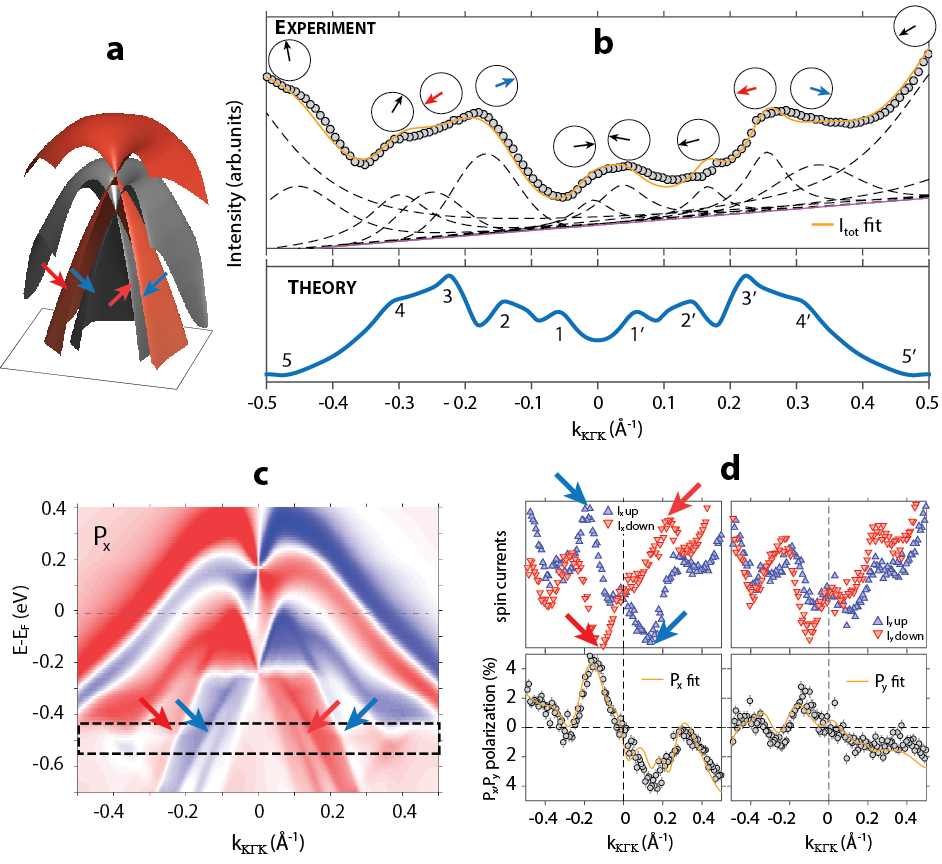}
\caption{(a) Simplified surface electronic structure model of the \aGeTe\ bulk band structure. (b) 3D vectorial spin analysis of data measured along \KGK\ at 0.5 eV binding energy. Theoretical momentum distribution curve (blue line) shows peaks \hbox{1-5} and \hbox{1'-5'}, all accounted for in the top panel to fit the total intensity (orange line). The arrows above each peak show the in-plane projection of the spin vector. (c) \Px-spin component of the semi-infinite band-structure calculations in \ref{F2}d. (d) (bottom) Measured  \Px\ and \Py\  spin polarization and fits (orange line), and (top) derived spin currents along the resolution-broadened energy range indicated by the dashed rectangle in (c).  The main Rashba-type bulk spin splitting is indicated with red/blue arrows in (c) and (d).}
\label{F3}
\end{figure}

Another approach to reveal the dispersive character of the SR-bands in \aGeTe\ is a \kz\nh dispersion movie in the \KGK\ mirror plane (see ancillary files). The scan stretches over two Z\nh points in the 3D Brillouin zone  and it shows that upon band-gap opening the SR-band separates from the bulk Rashba band and near the maximum gap at the $\Gamma$~point (\hn$\approx$400 eV) it disappears. As the gap is narrowing again in the \kz\nh scan, they reappear and disperse side-by-side with the bulk bands toward \EF\ such that at the Z-point they can be resolved only in a second derivative of the measured band map (Fig.~\ref{F1}b).    


From a technological point of view the pure surface Rashba bands $S_{1,2}$ and their resonances are less important because, as already mentioned, on capped \aGeTe\ surfaces they are completely quenched. Interestingly, their spectral signatures are also easily explored by variation of dipole transition matrix elements. As shown by Fig.~\ref{F2}f,g, the $p$ and $s$-polarized light in an experimental geometry depicted in Fig.~\ref{F2}e almost toggles on and off the bulk and bulk-derived bands.  This suggests that the dipole selection rules can be used to select the states originating in Ge and Te $p_z$-orbitals, oriented perpendicular to sample surface along the \ooo\ direction. This is also confirmed by one-step photoemission calculations on the bottom panels of Fig.~\ref{F2}(f,g), made by a fully relativistic one-step model in its spin-density matrix formulation as implemented in the SPR-KKR package \cite{Minar_RPP, JBraun_Rashba}. 

\begin{figure}[b]
\includegraphics[width=8cm]{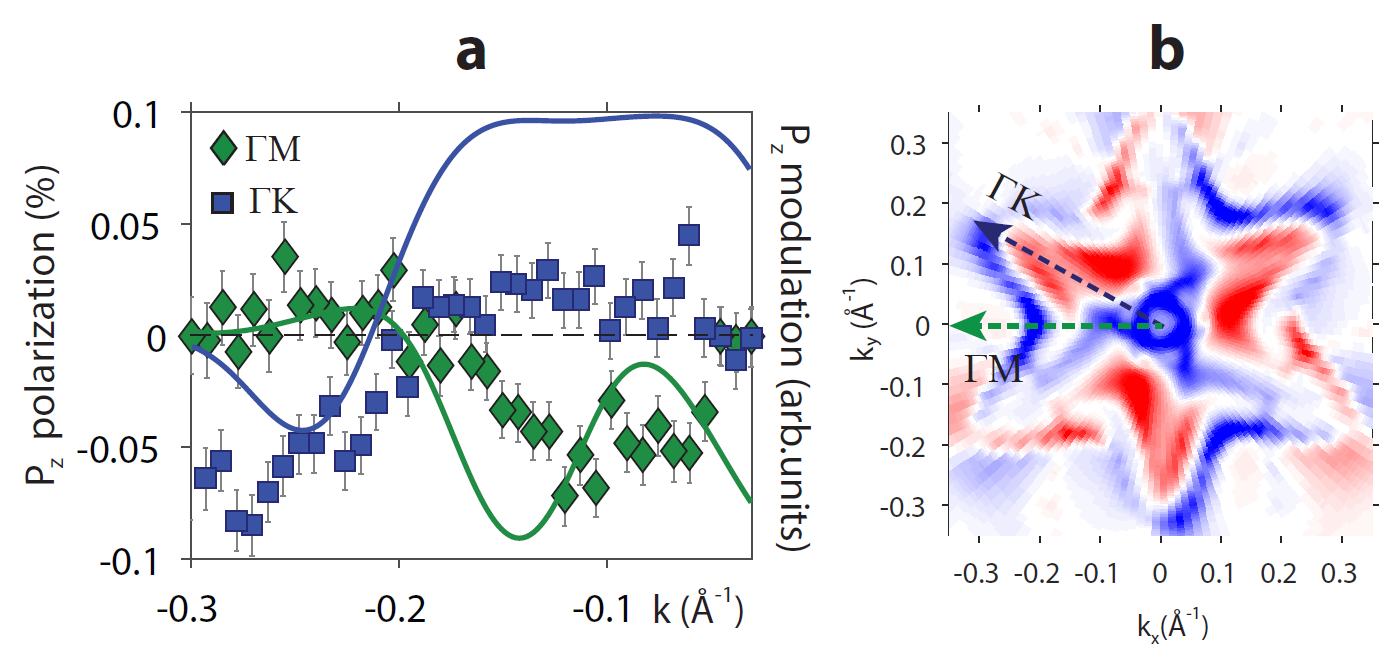}
\caption{(a) Out-of-plane spin-polarization \Pz\ measured of \aGeTe\ at 0.5 eV binding energy along \GM\ and \GK (symbols). Full lines show the corresponding \Pz\ modulations for the semi-infinite crystal calculations in (b).}
\label{F32}
\end{figure}

\begin{figure*}[t!]
\centering
\includegraphics[width=.8\textwidth]{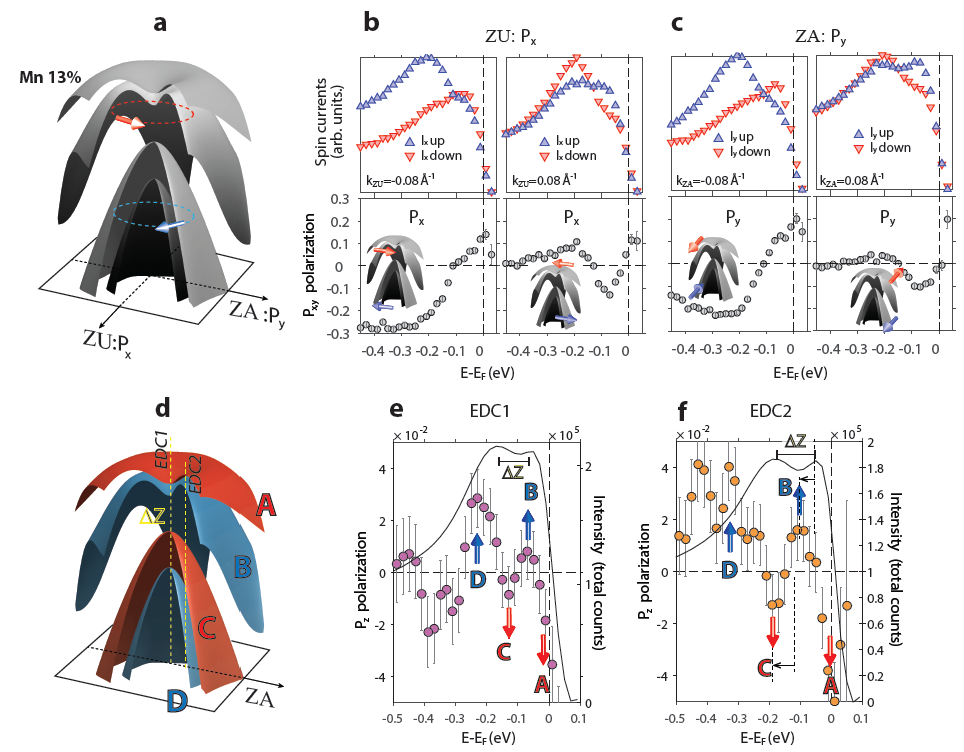}
\caption{(a) Simplified surface electronic structure model of the \GMT\ bulk band structure with in\nh plane spin texture above and below the Zeeman gap $\Delta$Z. (b,c) Measured \Pxy\ spin polarization (bottom) and corresponding spin currents (top) for off-normal emission ($\pm$0.08\kA) along \ZU\ and \ZA\ directions from \GMTD. (d) Simplified model of the out-of-plane spin texture \Pz\ from as\nh grown  \GMTD\ samples, deduced from data measured in normal (e) and off-normal (f) emission. The red/blue arrows indicate the spin texture, the horizontal arrow in (f) indicate the shift of the bands B-C which define the Zeeman gap of $\approx$100 meV, as observed in the total counts (solid line in e,f).}
\label{F4}
\end{figure*}  

For a practical description of the \aGeTe\ bulk electronic structure in surface-sensitive ARPES, Fig.~\ref{F2}(b,c) shows a simple cartoon view of the bulk and bulk-derived SR bands depicted in black and red, respectively. Until new detection schemes in SARPES become available in \hbox{soft-X} regime capable to investigate pure bulk states \cite{Strocov:2015,Fadley:2012}, \aGeTe\ SARPES data will always integrate the spectral intensity from both the SR and bulk bands, as seen in Fig.~\ref{F1}b. The band map clearly resolves the narrow-gapped bulk states (black dashed lines) and their surface resonance-replica (red dashed lines) screening the bulk states and shifted up to \EF. Visualizing SARPES data with energy distribution curve (EDC) as seen in Fig.~\ref{F1}b one should always keep in mind that the SR and bulk bands intersect these bands in four points denoted A-C for surface-resonances, and B-D for the bulk bands. 

\section{Experiments versus first-principles calculations}

To illustrate the validity of the  electronic structure model, we compare  rigorous first principles calculations to the experimental data. Figure~\ref{F3} summarizes SARPES data measured in the \KGK\ mirror plane around a binding energy of 0.5 eV along the momenta denoted in the dashed frame in panel (b). The calculations predict that the electronic structure is highly modulated with up to ten peaks labelled 1-5 and 1'-5', respectively. 
Also the measured spin texture is highly modulated, in our case the appearance of individual peaks is well accounted in both experiment and theory (top panel in Fig.~\ref{F3}b). The spin fitting is comprehensively described using a 3D vectorial analysis which fits MDC total intensity and measured 3D spin polarizations \cite{JK_PRB,JK_GMT,JK_opGeTe,Fabian_NJP} (orange lines in Fig.~\ref{F3}b,d). The obtained spin vectors from individual peaks, projected in the \{x,y\} plane (see experimental geometry in Fig.~\ref{F2}e) are shown in Fig.~\ref{F3}a. Consistent with the calculated spin-resolved band-map in panel (c), the main in-plane spin currents are detected along the {x}-direction. Along that direction there are two prominent spin currents highlighted by the red and blue arrows in panel (d). SARPES MDC maps these currents as two main bulk-like Rashba bands, which is evidenced by their antiparallel \Pxy\ spin vector alignment (red-blue arrows in panel b). We note that the spin-switching of these two bands was extensively tested in \textit{operando} SARPES in field effect devices to show that their manipulation by E-fields is possible \cite{JK_opGeTe}. These experimental observations give us confidence that the highly modulated spin texture can indeed be simplified as depicted in panel (a).

\begin{figure*}[t!]
\centering
\includegraphics[width=.8\textwidth]{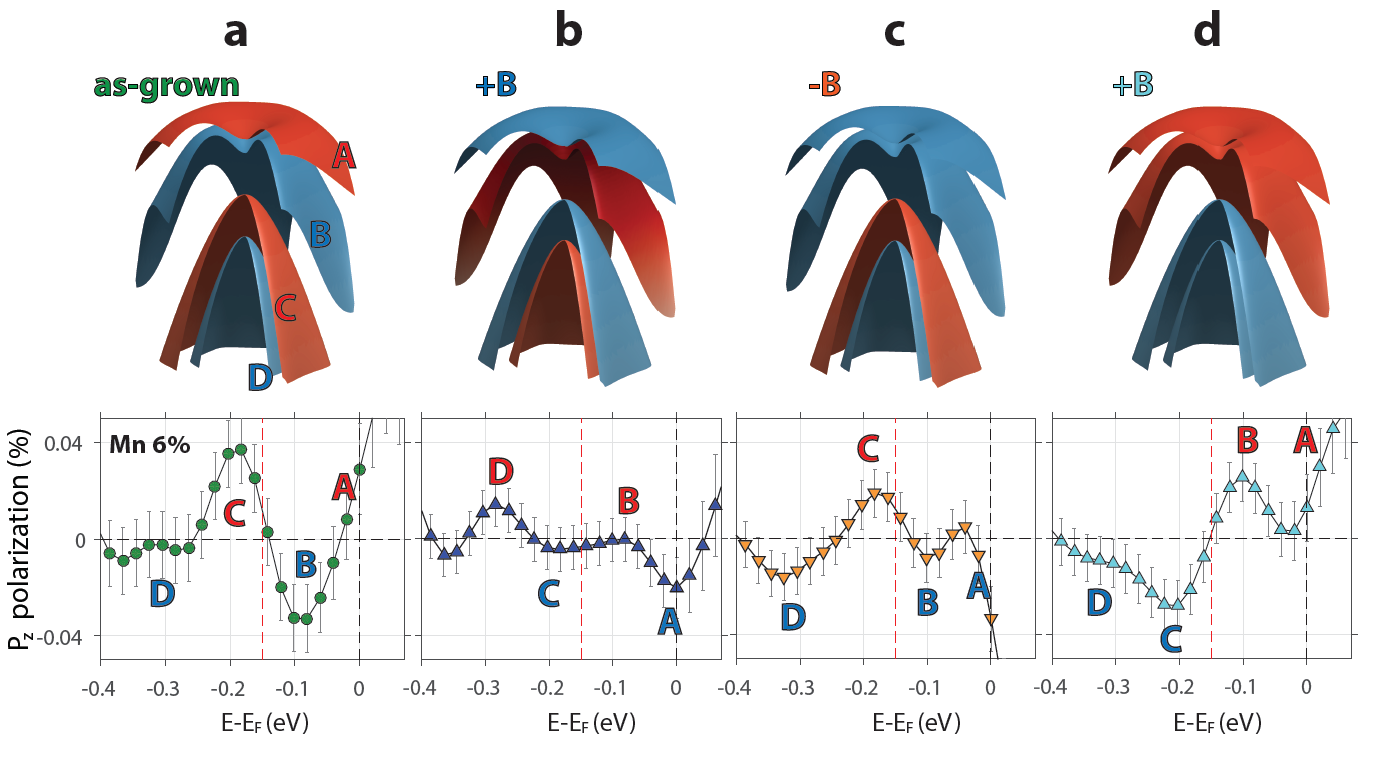}
\caption{(a-d) Deployment of the \Pz\ spin-texture of the bands A-D from \GMTDD\ measured in remanent magnetization after magnetizing the sample with $\pm$700 Gauss (see text for details).}
\label{F5}
\end{figure*}

Equally highly modulated is the out-of-plane spin-polarization \Pz\ measured at the same binding energy, shown in Fig.~\ref{F3}.  SARPES data in panel (a) is visualized as spin-resolved MDCs, which we relate to calculations in panel (b), measured along the \GM\ and \GK\ directions as denoted by blue/green arrows. The measured \Pz\ modulation shows excellent agreement with the first-principles calculations and confirms our detailed understanding of the \aGeTe\ \Pz\ warping around the Z-point, in agreement with our previous studies \cite{JK_PRB}.

\section{Modification of $\alpha$-G\MakeLowercase{e}T\MakeLowercase{e} by Mn-doping}

Figure~\ref{F4} visualizes SARPES data from \GMTD. Panels (a-c) summarize the in-plane \Pxy\ spin windings above and below the Zeeman gap, and panels (d-f) summarize the out-of-plane \Pz\  spin texture. For clarity the simplified surface electronic structure with the Rashba splitting and Zeeman gap is depicted in panel (a). The Zeeman gap opens up around a binding energy of 0.1 eV. In agreement with previous studies \cite{JK_GMT}, the gap size measured in total intensity is $\Delta_Z\approx$100 meV (Fig.~\ref{F4}e-f). The splitting of the surface electronic structure to B-D bulk and A-C bulk-derived bands becomes evident in the \Pz\ spin texture. In order to confirm the presence of the A-D bands already mentioned in Fig.~\ref{F1}b, data are measured in normal emission (Fig.\ref{F4}e) and off-normal emission (Fig.\ref{F4}f), respectively. The subtle shift in binding energy of the peaks B and C in panel (e) indicated by horizontal arrows in panel (f) is confirming dispersion of all the bands A-D and thus also the simplified electronic scheme of the bulk-derived bands in panels a,d.

\section{Field control of spin texture}

In \aGeTe\, electric-field control of the spin windings is possible by deposition of a metallic gate electrode on the surface. 
Applying a gate voltage induces a change in the spin polarization, however, we find that the endurance of the spin switching caused by changing the  field direction is  limited due to unipolar FE fatigue and other effects such as FE domain pinning \cite{JK_opGeTe}.  Moreover, epitaxial \aGeTe\ films typically display a multidomain structure \cite{Springholz_PRL,JK_PRB,Calarco_GeTe} in which polarization reversal may involve intermediate steps via oblique domains rather than direct switching along the \ooo\ axis which is coupled to the a full spin texture reversal \cite{JK_opGeTe}.

\GMT\ appears to have a weaker pinning of the FE polarization because the off-center displacement of the Te atom with respect to the Ge atoms in \GMT\ decreases with increasing Mn content \cite{Springholz_PRL,Kriegner_2016}. This reduces the energy barriers for switching of the atomic positions in the FE reorientation and thus leads to a softening of the FE properties while simultaneously acquiring magnetoelectric properties. Thus, from an application point of view, \GMT\ fulfills all criteria for mutual control of magnetism and ferroelectricity via magnetoelectric coupling effects, which is a unique material property \cite{Eerenstein2006,Fiebig2005}.
 
In order to emphasize the close relation between \aGeTe\ and \GMT, Fig.~\ref{F5} summarizes the B-field control of \GMTDD\ in which the size of the Zeeman gap is less than 50 meV \cite{JK_GMT}. Data were measured at the CASSIOP{\`E}E beamline at the Soleil synchrotron in remanent magnetization and show how the \Pz\ spin-texture from as-grown samples develops in consecutive sample magnetization cycles. Contrary to the E-field manipulation of the \aGeTe\ spin texture in which the spin control is basically stalled after the second cycling, the B-field control from \GMTDD\ is found to change after each sample magnetization. We note that after the third magnetization cycle (Fig.~\ref{F5}d) the \Pz\ spin texture stabilizes in a configuration as predicted by theory for bulk \GMT\ \cite{JK_GMT, JK_opGeTe}. 

SARPES data in Fig.~\ref{F4}-\ref{F5} confirm that apart from the Zeeman gap, the simplified surface electronic structure model of  \aGeTe\ also applies for \GMT\ and that manipulating the spin-texture by external fields in photoemission impart additional degrees of freedom associated with surface resonances, as seen in the gradual deployment of the \Pz\ spin-texture in Fig.~\ref{F5}.  
}
Our experimental observations suggest that there are certain volatile degrees of freedom in the surface electronic structure and in the spin-texture which give rise to complex switching paths. Consequently they may result in unconventional spin texture evolutions upon manipulation by external fields \cite{JK_GMT, JK_opGeTe} or by tuning the \aGeTe\ surface termination. For example the energetically less favorable \aGeTe\ surface termination with Ge-atoms discussed in Ref.~\citen{Bertacco_GeTe}  according to the simplified surface electronic structure affects only the top-most surface-resonance sheet A sitting right at \EF, rather than a the full switching which extends to the bulk Rashba bands. 

\section{Conclusions}

By comprehensive (S)ARPES mapping of the electronic structure we have evaluated in detail the spin-resolved electronic structure of the ferroelectric and multiferroic Rashba semiconductors \aGeTe\ and \GMT\. The strong spin-orbit effect entails large spin splitting of the surface electronic structure consisting of surface and surface resonant states, which are screening the bulk Rashba bands. The different contributions can be separated and analyzed by combining measurements at different photon energies and photon polarizations. Independently of the substrates (BaF$_2$, InP or Si [111]) used for thin film deposition, ARPES consistently indicates that all the band types possess their own Dirac point and that the surface states have the Dirac point in the unoccupied states. Our experimental finding are in excellent agreement with \textit{ab-initio} calculations based on the multiple scattering approach, density functional theory and semi-infinite crystal calculations with included spin-orbit coupling, as described in Ref.~\citen{JK_PRB, JK_GMT}. This leads to a simplified model of the \aGeTe\ and \GMT\ surface electronic structure, its validity is shown by comparison to rigorous \textit{ab\nh initio} calculations. Our experimental results confirm the the coupling between the ferromagnetic and ferroelectric order in \GMT\ because the Rashba-type spin texture is clearly influenced by the magnetization switching. This is the main precondition for functional spintronic applications,  but presently the magneto-electric coupling imposes limited functionality due to the complex switching paths of the Rashba spin textures  even at temperatures around 35~K. 

\section{Acknowledgements}  

This work was supported by the Swiss National Science Foundation Project No. PP00P2\_144742\/1 and n.200021 146890.  G.S. and V.V.V. acknowledge support from the Austrian Science Funds (SFB-025, IRON). JM would like to thank CEDAMNF project (CZ.02.1.01/0.0/0.0/15\_003/0000358) funded by the Ministry of Education, Youth and Sports of Czech Republic.

\bibliographystyle{apsrev4-1}
%

\end{document}